\documentclass[a4paper]{article}

\usepackage{arxiv}
\usepackage[utf8]{inputenc} % allow utf-8 input
\usepackage[T1]{fontenc}    % use 8-bit T1 fonts
\usepackage[hidelinks]{hyperref}       % hyperlinks
\usepackage{url}            % simple URL typesetting
\usepackage{booktabs}       % professional-quality tables
\usepackage{amsfonts}       % blackboard math symbols
\usepackage{nicefrac}       % compact symbols for 1/2, etc.
\usepackage{microtype}      % microtypography
\usepackage{lipsum}
\usepackage{graphicx}
\usepackage{array}
\usepackage{appendix}
\usepackage{amsmath}
\usepackage{balance}
\usepackage{authblk}
\usepackage{lipsum}

\DeclareMathOperator{\EX}{\mathbb{E}} % expected value

\newcommand\blfootnote[1]{%
  \begingroup
  \renewcommand\thefootnote{}\footnote{#1}%
  \addtocounter{footnote}{-1}%
  \endgroup
}

\title{Efficient acoustic feature transformation in mismatched environments using a Guided-GAN}

\author[a,b]{
  Walter Heymans
}
\author[a,b,c]{
Marelie H. Davel
}

\author[d]{
Charl van Heerden
}

\affil[a]{
  Faculty of Engineering,
  North-West University, South Africa
  
    \url{https://engineering.nwu.ac.za/must}
}
\affil[b]{
Centre for Artificial Intelligence Research (CAIR), South Africa}
\affil[c]{
  National Institute for Theoretical and Computational Sciences (NITheCS), South Africa
}
\affil[d]{
  Saigen, South Africa
}

\begin{document}
\maketitle
\blfootnote{This is a preprint - the final authenticated publication is available online at:\\ \url{https://doi.org/10.1016/j.specom.2022.07.002} \\
© 2022. This manuscript version is made available under the CC-BY-NC-ND 4.0 license \url{https://creativecommons.org/licenses/by-nc-nd/4.0}}

\begin{abstract}
We propose a new framework to improve automatic speech recognition (ASR) systems in resource-scarce environments using a generative adversarial network (GAN) operating on acoustic input features. The GAN is used to enhance the features of mismatched data prior to decoding, or can optionally be used to fine-tune the acoustic model. We achieve improvements that are comparable to multi-style training (MTR), but at a lower computational cost. With less than one hour of data, an ASR system trained on good quality data, and evaluated on mismatched audio is improved by between 11.5\% and 19.7\% relative word error rate (WER). Experiments demonstrate that the framework can be very useful in under-resourced environments where training data and computational resources are limited. The GAN does not require parallel training data, because it utilises a baseline acoustic model to provide an additional loss term that guides the generator to create acoustic features that are better classified by the baseline. 

\end{abstract}
environments\blfootnote{\textit{E-mail addresses}: 
\href{mailto:walterheymans07@gmail.com}{walterheymans07@gmail.com} (W. Heymans), \href{mailto:marelie.davel@nwu.ac.za}{marelie.davel@nwu.ac.za} (M.H. Davel), \href{mailto:charl@saigen.com}{charl@saigen.com} (C. van Heerden),}

\section{Introduction}
Automatic speech recognition (ASR) systems generally struggle to perform well when there is a mismatch between the data that was used to train the ASR system and new data that is decoded. This is especially true in under-resourced environments, where it is often prohibitively expensive to collect enough transcribed training data to cover even the most common application environments. 

For some under-resourced languages, there are ASR systems that achieve low word errors rates (WERs) on audio similar to what the systems were trained with. However, when such a system is used to decode mismatched data, the WER is significantly higher. Some of the challenging aspects that lead to mismatched conditions include background noise, reverberation, encoding artifacts and microphone distortion. In cases where it is possible to collect a small dataset of training data that better matches the test conditions, it is usually very expensive to retrain an entire system to include the new data. This effect is magnified when new mismatched data regularly needs to be decoded. When computational resources are limited, it is not feasible to regularly retrain models or have multiple models that run in parallel.

In order to be useful in many different, constantly changing environments, a good baseline ASR system requires an efficient technique to adapt to the changing environments. An attractive solution is feature compensation because you can replace the adaptation technique without changing the acoustic model~\cite{deng2011front}. The features of noisy audio are changed or enhanced to increase the accuracy of the acoustic model~\cite{li2017improved}. Methods that are used as a front-end processing technique include masking time and frequency components~\cite{narayanan2014investigation}, speech enhancement~\cite{weninger2015speech}, deep denoising autoencoders~\cite{feng2014speech} and generative adversarial network (GAN) based feature enhancement~\cite{sriram2018robust, donahue2018exploring}. An alternative to feature compensation is model compensation, where the acoustic model is adapted in some way. Model compensation usually provides better performance gains than feature compensation, but at a much higher computational cost~\cite{deng2011front}.

Multi-style training (MTR) is one of the most popular and well-established techniques used to address mismatch by allowing the acoustic model to learn robust representations of the data. MTR aims to transform the training data to be more representative of the testing data. New training datasets are created from an existing set by adding a series of MTR styles. These can include changing the speed and volume~\cite{ko2015audio}, speech style~\cite{lippmann1987multi} or sampling rate~\cite{li2012improving}, adding time and frequency distortions~\cite{park2019specaugment} or background noise, and simulating reverberation~\cite{szoke2019building}. The styles are typically chosen without knowledge of the testing conditions and must still be able to handle a wide variety of mismatch. In addition, the number of styles that are added must be considered because the computational cost of training an ASR system increases significantly with each style that is added.

Recently, a number of GAN-based feature enhancement techniques have been proposed to improve ASR robustness to noise and reverberation~\cite{sriram2018robust, donahue2018exploring, wang2018investigating}. A GAN framework uses two networks that compete in a min-max game and improve by learning from each other~\cite{goodfellow2014generative}. A generator network learns to create new samples using random noise as input. Another network, called the discriminator, is used to determine if the sample came from the true data distribution or the generator. The output of the discriminator is maximised for real samples and minimised for generated samples. The generator is trained using the output from the discriminator to improve the quality of the samples it generates. In ASR, the generator uses speech features or embeddings as input and aims to produce new features that are more robust to noise or reverberation~\cite{sriram2018robust, donahue2018exploring}. GAN-based feature enhancement techniques can achieve very good performance improvements when a clean trained baseline ASR system is used~\cite{sriram2018robust, donahue2018exploring, wang2018investigating}.

In this work, we investigate the use of GANs to transform acoustic features of mismatched audio in under-resourced environments. We want to improve the performance on a new mismatched test set with a limited computational budget, provided that an ASR system trained on good quality data already exists. Our GAN is trained to transform acoustic features of mismatched audio to be better classified by the acoustic model. Our technique can be applied to most acoustic features including Mel-frequency cepstral coefficients (MFCCs), filter-banks and speaker-adapted transforms like feature-space maximum likelihood linear regression (fMLLR). For the purposes of this paper, we focus specifically on WAV49-encoded audio \cite{etsi6300, van2019asterisk}, a full-rate GSM codec with a compression ratio of 10:1, often used by South African call centres to store large volumes of telephone calls. Call centre speech analytics is an important application of ASR, and due to the large volumes of audio generated by call centres, some form of compression is often encountered. This is why we specifically focus on noisy speech (also typical of call centre data) encoded with WAV49.

\section{Related work}
GANs can improve both deep neural network hidden Markov model (DNN-HMM)~\cite{wang2018investigating} and end-to-end ASR systems~\cite{sriram2018robust, donahue2018exploring} by creating indistinguishable representations between clean and noisy audio samples. A clean corpus is augmented by adding noise or room impulse responses to create a new noisy corpus. GANs used for DNN-HMM ASR systems normally operate on acoustic features (log-power spectra, MFCCs)~\cite{wang2018investigating}, and on log-Mel filterbank spectra or embeddings when used for end-to-end systems ~\cite{sriram2018robust, donahue2018exploring}. The generator maps the noisy features to the same representation as the corresponding clean features. The discriminator provides feedback to the generator using the clean and noisy features as input. An L1 or mean squared error loss between clean and noisy samples is minimised which assists the generator with the mapping~\cite{sriram2018robust, donahue2018exploring}. This only works when the noisy set is created from the clean set, because the loss requires both samples to be of the same frames in a recording. All these approaches rely on the assumption that you know what the test conditions are (noise or reverberation). They are only as effective as the conditions you can create using a clean dataset. Still, GANs such as these that are used to improve features or embeddings have been shown to outperform data augmentation and DNN-based enhancement techniques~\cite{sriram2018robust, wang2018investigating}.

These approaches are designed primarily for noise robustness and dereverberation, not necessarily mismatched audio. It is extremely difficult to accurately produce a noisy dataset using only clean audio if the conditions of the testing data are unknown. Sometimes it is possible to collect a small training dataset with matched conditions to the new test set. In this scenario, the previous GAN-based enhancement techniques cannot be used, because a parallel training set (clean version) does not exist.

An approach different from feature enhancement is to use a GAN to fine-tune an end-to-end ASR system~\cite{haidar2021fine}. The entire ASR model is treated as a generator network. The ASR system is trained normally until convergence occurs. After training, a discriminator network is tasked to determine if the transcriptions came from the ASR system or are ground truth. Since the ASR model is already trained, the output is very similar to the ground truth, which makes the task very difficult for the discriminator. The GAN can improve output transcriptions of an ASR model that has already converged. Adding the GAN-based fine-tuning technique yields consistent WER improvements~\cite{haidar2021fine}.

Previous work focused on noisy and reverberant speech, with little attention paid to compressed audio. None of the GAN-based enhancement techniques are designed to work with a newly transcribed dataset that does not have a corresponding clean version. We are not aware of any research that utilises GANs for acoustic feature transformation in under-resourced ASR or call centre environments.

\section{Guided-GAN}
We use a GAN to improve an ASR system on mismatched (noisy and compressed) audio. Our approach can use any GAN loss function, but changes the loss in order to utilise an existing ASR system to guide the GAN training process. We refer to this architecture as a Guided-GAN.

\begin{figure}[b]
  \centering
  \includegraphics[width=0.7\linewidth]{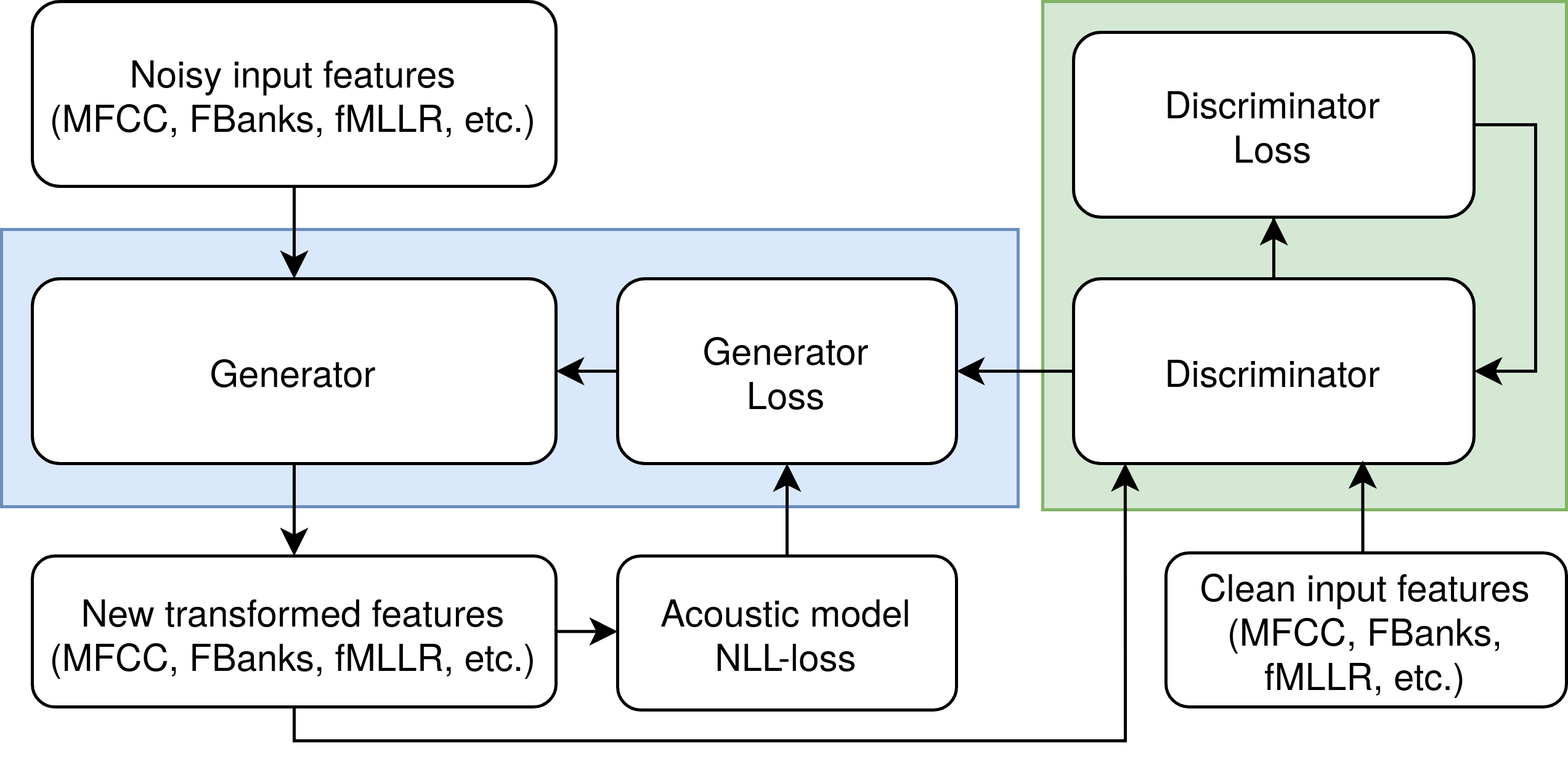}
    \caption{Diagram of a Guided-GAN training process.}
    \label{fig:am_gan}
\end{figure}

Two datasets are used during GAN training. The first set can be a clean dataset or the same data used to train the acoustic model (known conditions). The second set, which is used as input for the generator, is a noisy or mismatched dataset that does not come from the same distribution as the clean set. Figure \ref{fig:am_gan} shows a diagram of the Guided-GAN training process. Acoustic features of the mismatched audio are given to the generator as input. The generator then creates a new vector of the same dimension, which is used as input to both a pre-trained DNN acoustic model and the discriminator. The negative log-likelihood (NLL) loss of the acoustic model is added to the generator's loss to assist it in generating realistic samples of the correct class. The input to the discriminator is clean (or have known conditions) and generated features for which the output is maximised and minimised respectively. The goal of the GAN is to reduce the performance difference between data with known conditions and a new mismatched test set, for which a small transcribed training set is available. When the system is used to evaluate a test set, only the generator network is used to transform the noisy features before handing them over to the downstream acoustic model.

\subsection{Loss functions}
\label{sec:loss_functions}
Any GAN loss function can be extended by adding the guiding term to the generator's loss. We use three different loss functions and select the one that is best suited for our approach. The first is an adapted version of a spectral normalisation GAN (SN-GAN)~\cite{miyato2018spectral} in which we extend the generator loss function with an additional term. Specifically, we use the labels of the noisy sample to calculate the NLL loss of the cleaned sample. This value is added to the loss function to guide the generator to create features that are more likely to be classified correctly by the target acoustic model. An SN-GAN uses the same loss functions as a Wasserstein GAN (WGAN), but uses spectral normalisation instead of weight clipping in the discriminator to satisfy the Lipschitz constraint~\cite{miyato2018spectral}. The standard WGAN loss for the generator is given by:

\begin{equation}
    \mathcal{L}_G = -\EX_{\mathbf{\hat{x}} \sim p_g}[D(\mathbf{\hat{x}})],
    \label{eq:sn_wgan_loss}
\end{equation}

\noindent where $p_g$ is the model distribution implicitly defined by $\mathbf{\hat{x}} = G(\mathbf{\Tilde{x}})$ with $\mathbf{\Tilde{x}}$ the noisy input features, and $D(\mathbf{\hat{x}})$ is the output of the discriminator that can be a positive or negative value. We add the NLL loss to Eq. (\ref{eq:sn_wgan_loss}):

\begin{equation}
    \mathcal{L}_G = -\EX_{\mathbf{\hat{x}} \sim p_g}[D(\mathbf{\hat{x}})] 
    - \lambda \cdot \EX_{\mathbf{\hat{x}},\Tilde{y} \sim p_{data}}\log{ p_{am}(\Tilde{y}|\mathbf{\hat{x}})},
    \label{eq:guided_gan_loss}
\end{equation}

\noindent where $\lambda$ is a hyperparameter, $p_{am}$ is the probability defined by the trained acoustic model, and $\Tilde{y}$ is the senone class label of the noisy features. The WGAN loss function for the discriminator is given by:

\begin{equation}
\begin{split}
    \mathcal{L}_D = -\EX_{\mathbf{x} \sim p_d}[D(\mathbf{x})] + \EX_{\mathbf{\hat{x}} \sim p_g}[D(\mathbf{\hat{x}})]
    \label{eq:sn_wgan_loss_d}
\end{split}
\end{equation}

\noindent where $p_d$ is the real distribution defined by the clean samples $\mathbf{x}$. Spectral normalisation is used in the discriminator to ensure that the Lipschitz constraint is satisfied \cite{miyato2018spectral}. We also compare the SN-GAN loss functions to the standard non-saturating GAN (NS-GAN)~\cite{goodfellow2014generative} and the WGAN with gradient penalty (WGAN-GP)~\cite{gulrajani2017improved}. The NS-GAN loss function we use for the generator is:
\begin{equation}
\begin{split}
    \mathcal{L}_G^{NS-GAN} = - \mathbb{E}_{\mathbf{\hat{x}} \sim p_g}[\log{D(\mathbf{\hat{x}})}] - \lambda \cdot \EX_{\mathbf{\hat{x}},\Tilde{y} \sim p_{data}}\log{ p_{am}(\Tilde{y}|\mathbf{\hat{x}})}
    \label{eq:nsgan_g_loss}
\end{split}
\end{equation}
with a corresponding discriminator loss function:
\begin{equation}
\begin{split}
    \mathcal{L}_D^{NS-GAN} = -\mathbb{E}_{\mathbf{x} \sim p_d}[\log{(D(\mathbf{x}))}] - \mathbb{E}_{\mathbf{\hat{x}} \sim p_g}[\log{(1 - D(\mathbf{\hat{x}}))}].
    \label{eq:nsgan_dloss}
\end{split}
\end{equation}

The WGAN-GP loss function for the generator is the same as for the SN-GAN in Eq. \ref{eq:sn_wgan_loss}. The Lipschitz constraint is satisfied in the discriminator by adding a gradient penalty term to Eq. \ref{eq:sn_wgan_loss_d}:
\begin{equation}
\begin{split}
    \mathcal{L}_D^{WGAN-GP} = -\mathbb{E}_{\mathbf{x} \sim p_d}[D(\mathbf{x})] + \mathbb{E}_{\mathbf{\hat{x}} \sim p_g}[D(\mathbf{\hat{x}})] + \\ \lambda_{gp}\mathbb{E}_{\mathbf{\hat{x}} \sim p_g}[(||\nabla D(\alpha\mathbf{x} + (1 - \alpha)\mathbf{\hat{x}})||_2 - 1)^2],
    \label{eq:wgan_gp_d}
\end{split}
\end{equation}
where $\lambda_{gp}$ is a hyperparameter controlling the gradient penalty, and $\alpha$ is a uniform random number between zero and one.

\subsection{Network architectures}
\label{sec:network_architectures}
We compare two different network architectures for the task. The first architecture is based on the encoder-decoder architecture of a speech enhancement GAN (SEGAN)~\cite{pascual2017segan}. The original network operated on raw audio and added random noise between the encoder and decoder networks. Our implementation uses fMLLR features as input and we do not add any noise. This architecture is described in Section \ref{sec:encoder_decoder}.

The second architecture we use improves the efficiency of the first by reducing the number of layers and removing the downsampling and upsampling processes. The parameters are fewer and allows for much faster training. This architecture is referred to as a fully-convolutional network because the output dimension will always equal the input dimension for any given input dimension. We later show in Section \ref{sec:clean_wav49_librispeech} that the performance of the two networks is similar in terms of the final WER for a system. The fully-convolutional network architecture is described in Section \ref{sec:fully-convolutional}.

\subsubsection{Encoder-decoder architecture}
\label{sec:encoder_decoder}
\begin{figure}[hbt]
  \centering
  \includegraphics[width=0.7\linewidth]{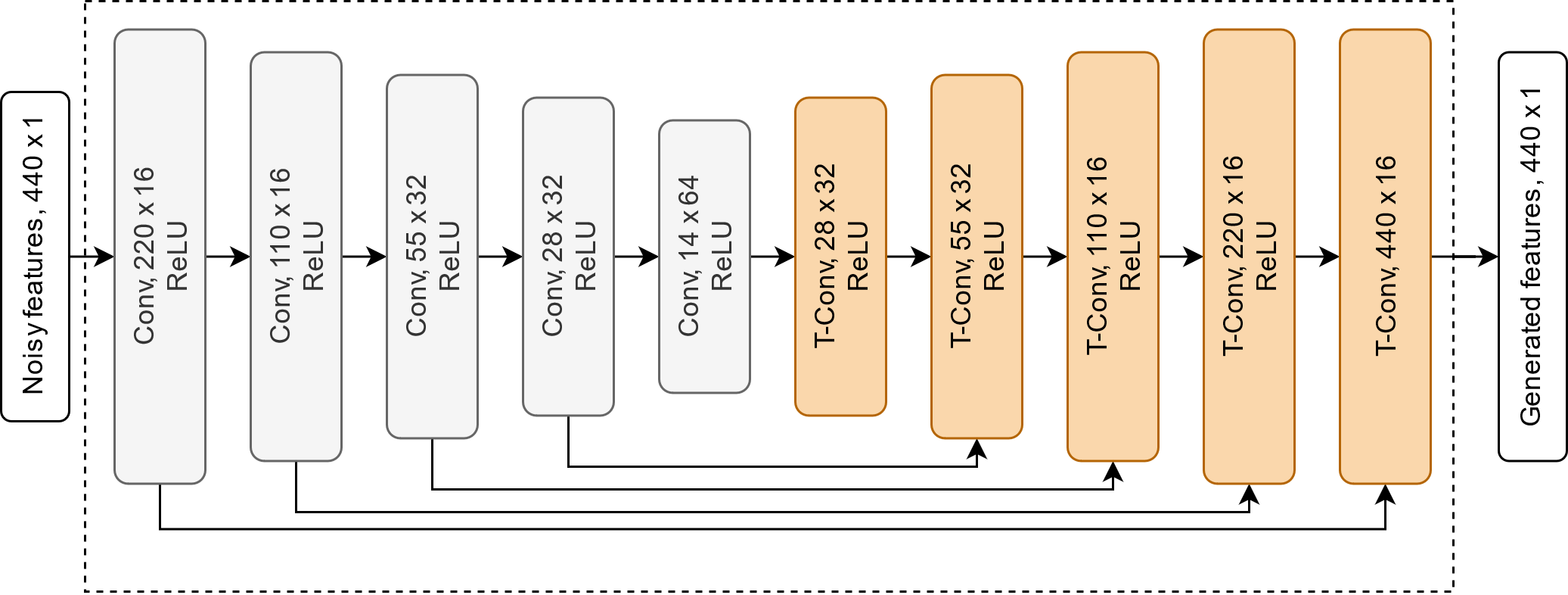}
    \caption{Diagram of encoder-decoder generator network architecture. (`Conv' is a convolutional layer with a stride of 2. `T-Conv' is a transposed convolutional layer.)}
    \label{fig:generator_unet}
\end{figure}

\begin{figure}[ht]
  \centering
  \includegraphics[width=0.75\linewidth]{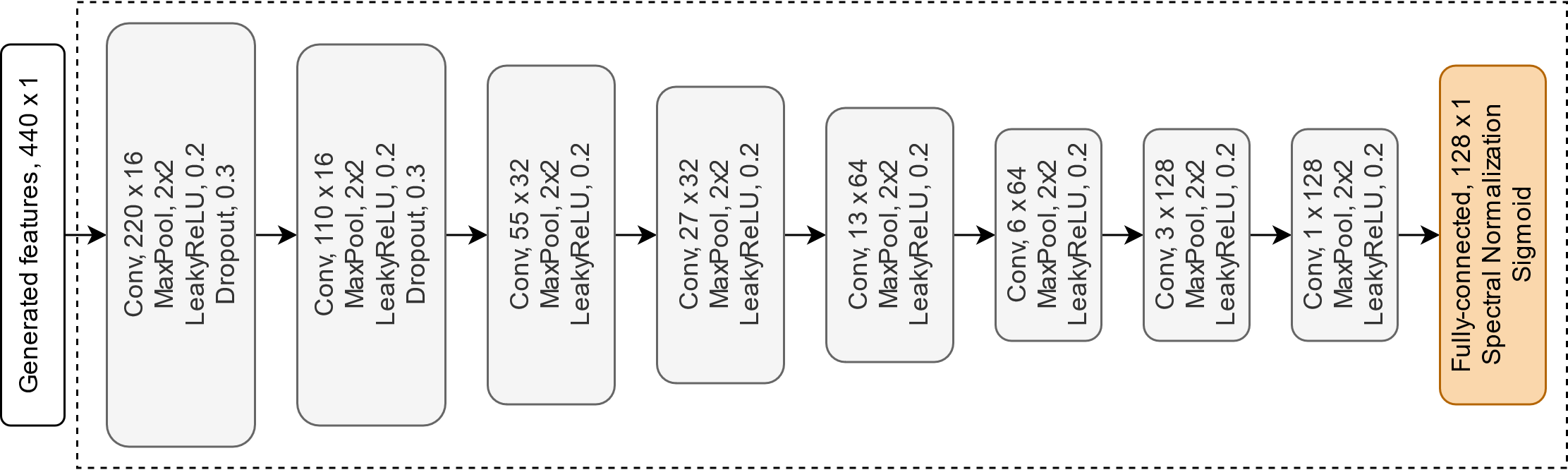}
    \caption{Diagram of discriminator network architecture for the encoder-decoder generator. (`Conv' is a convolutional layer.)}
    \label{fig:discriminator_large}
\end{figure}

A diagram of the generator and discriminator architectures for the SEGAN inspired networks are shown in Figures \ref{fig:generator_unet} and \ref{fig:discriminator_large}. We do not add any random noise or dropout to the generator during training or evaluation, which allows for a deterministic network. We found that the results were more consistent this way. The encoder network (shown in grey blocks) uses five convolutional layers with a stride of two for downsampling. The first two layers have a kernel size of seven, then two layers with a kernel size of five and the last layer has a kernel size of three. All layers in the generator, except for the output layer, use ReLU activation functions. The decoder network (shown in orange blocks) has five transposed convolutional layers with a stride of two for upsampling. The first and second layer uses a kernel size of four and five, respectively, followed by three layers with a kernel size of six. The choices for the kernel sizes were based on experimental results and architectural reasons (to upsample to the same dimension as the corresponding layer in the encoder). Skip connections are used to splice the feature maps of the encoder layers to the corresponding decoder layers. This is done because the encoder network extracts high-dimensional features and would struggle to produce an accurate output if it only upsamples these high-dimensional features. By splicing the feature maps of the encoder network, the features before the encoder operated on them can assist the generator to produce a better output. The high-dimensional features then assist the layers in the decoder to improve the upsampling process by removing noise from the original features.

The discriminator network that works with the SEGAN-inspired generator has eight convolutional layers followed by a fully-connected output layer. The first layer uses a kernel size of 41, the rest of the layers use a kernel size of 13. The network downsamples the features with max-pooling layers. All convolutional layers use leaky ReLU activation functions with a negative slope of 0.2. Dropout is used only in the first two layers with a probability of 0.3. The fully-connected output layer uses spectral normalisation and has a sigmoid activation function.

\subsubsection{Fully-convolutional architecture}
\label{sec:fully-convolutional}
\begin{figure}[hb!]
  \centering
  \includegraphics[width=0.6\linewidth]{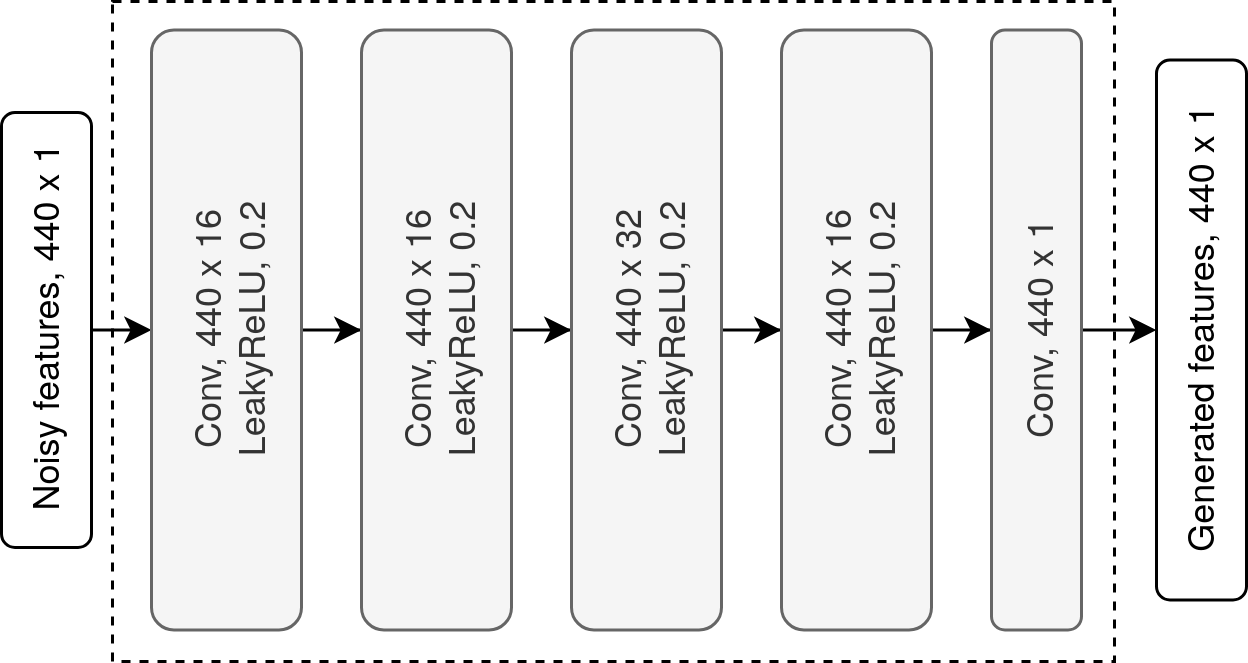}
    \caption{Diagram of fully-convolutional generator network architecture. (`Conv' is a convolutional layer.)}
    \label{fig:generator}
\end{figure}

The fully-convolutional generator architecture is shown in Figure \ref{fig:generator}. We use a kernel size of five for all layers, with zero padding to ensure that the dimension stays the same. The first four layers use leaky ReLU activation functions with a negative slope of 0.2. The last layer creates the final output without any activation function. We do not use dropout or add random noise at any stage in the generator.

\begin{figure}[hb]
  \centering
  \includegraphics[width=0.6\linewidth]{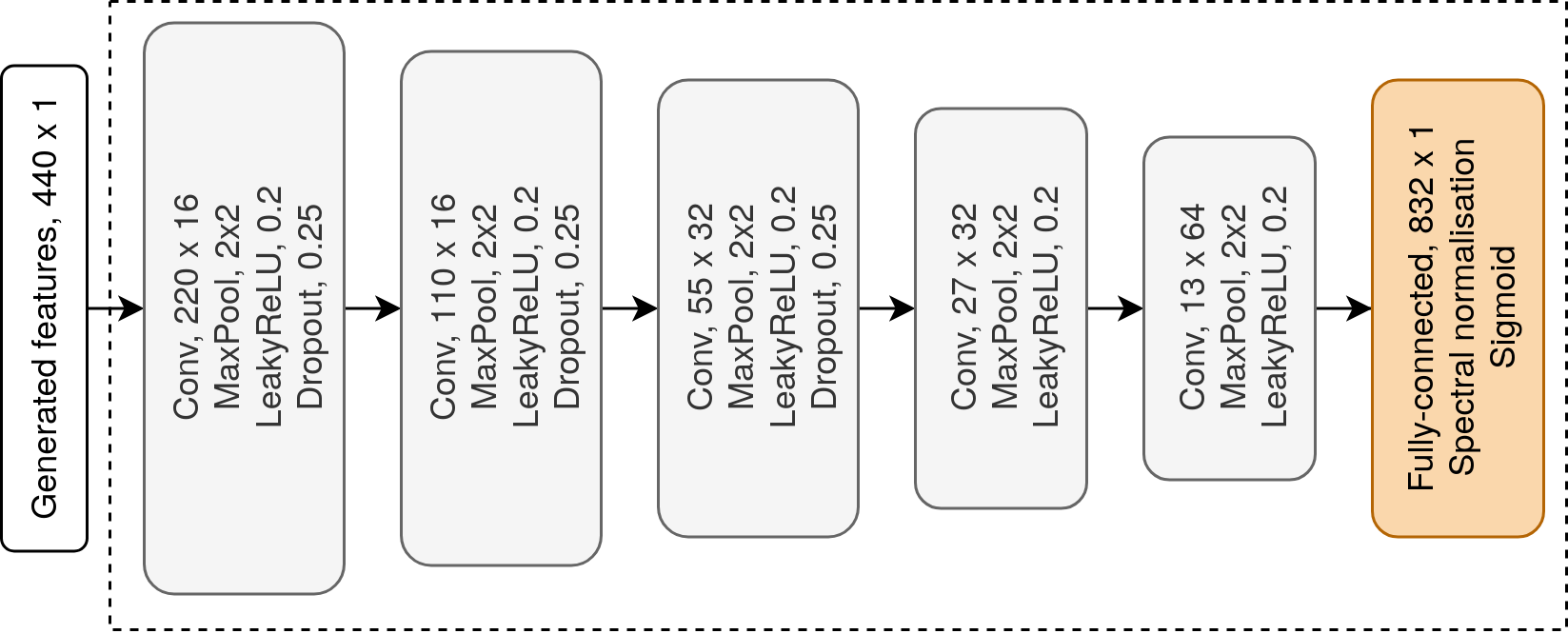}
    \caption{Diagram of fully-convolutional discriminator network architecture. (`Conv' is a convolutional layer.)}
    \label{fig:discriminator}
\end{figure}

The discriminator, shown in Figure \ref{fig:discriminator}, is a deep convolutional neural network with max-pooling layers and a fully-connected output layer. All layers, except the output, use leaky ReLU activation functions with a negative slope of 0.2. Dropout is used in the first three layers with a probability of 0.25. The output layer is a fully-connected layer with spectral normalisation and a sigmoid activation function. The sigmoid function is used to bound the output of the discriminator to ensure that the magnitude of the NLL loss and discriminator loss are comparable (ratio is controlled by the hyperparameter $\lambda$).

\subsection{Performance metric}
We use two different metrics for hyperparameter selection and reporting system performance. The most accurate and objective metric to select the best hyperparameters for the GAN is the WER on a development set. Measuring the WER requires decoding, which takes significant time and resources. An approach that is computationally much cheaper and correlates well with WER (correlation is shown in Appendix \ref{apx:correlation}), is to measure the senone error rate (SeER) of the baseline DNN acoustic model when given the generated features as input. SeER is measured before decoding, and is the error rate of tied context-dependent states. The hyperparameters of the GAN are tuned to get the lowest SeER on the development set. We also use early stopping to select the generator which had the lowest SeER. To evaluate and compare the final system performance of different models, we use WER.

\section{Experimental setup}
In this section, we describe the experimental setup that we use. We first describe the acoustic model and how it is trained, followed by the training and optimisation procedure for the GAN. We then describe our MTR setup that is used as a competitor for the GAN.  Finally, we describe how a controlled environment is created using the LibriSpeech corpus~\cite{panayotov2015librispeech}. The toolkit used for experimentation is publicly available\footnote{\url{https://github.com/walterheymans/pytorch-kaldi-gan}}.

\subsection{Acoustic model}
\label{sec:acoustic_model}
We use the baseline DNN acoustic model architecture of the Pytorch-Kaldi toolkit for our experiments~\cite{ravanelli2019pytorch}. The model is a 5 hidden layer multi-layer perceptron (MLP) with 1 024 units per layer; it uses ReLU activation functions, dropout and batch normalisation. We use fMLLR transformed input features with a context window of 11 frames. The network is trained with the SGD optimiser with a learning rate scheduler that halves the learning rate when the relative improvement\footnote{SeER is measured after each training epoch.} on the development set is less than 0.001. All networks are trained until convergence. The hyperparameters are optimised to minimise WER on the development set. The best batch size and learning rate combination is determined by performing a grid search. When scoring the development set, we select the best combination of language-model weight and word insertion penalty. We report WER on a held-out test set using the hyperparameters determined on the development set. On the standard LibriSpeech 100-hour training set~\cite{panayotov2015librispeech}, our results (9.18\% WER on \textit{test-clean}) are comparable with state-of-the-art results (using the 100-hour training set and the small language model, \textit{tg-small}) \cite{ravanelli2019pytorch, panayotov2015librispeech, kaldigitrepo}.

\subsection{GAN training and optimisation}
\label{sec:gan_training}
The GAN uses the same fMLLR features that is given to the acoustic model as input for both the generator and discriminator networks. The GAN is trained until the SeER converges on the development set. In initial experiments, the batch size and learning rates (independent learning rates for generator and discriminator) had the biggest impact on the SeER. We find the best combination of batch size and learning rates in a grid search. After the search is complete, we then expand the search in that area to find better results when varying the acoustic model $\lambda$. When changing the $\lambda$, we also change the batch size and learning rates again. We use the Adam optimiser with equal updates to the generator and discriminator. After the GAN is trained, the DNN acoustic model is evaluated with the generator operating on the features. In a second step, we also measure how fine-tuning the acoustic model can further reduce the WER. During fine-tuning, the pre-trained model is trained further using the cleaned features for a few iterations, until the SeER converges.

\subsection{Multi-style training}
\label{sec:mtr}
An MTR model is trained to serve as a competitor for the GAN model. We apply speed and volume perturbation to the datasets to create a more diverse training set. The perturbations are applied to the same datasets we use as `noisy' set for GAN training each time to ensure a fair comparison.

The speed and volume of the training data are changed with 10\% and 20\%, respectively~\cite{heymans2021mtr}. The speed is either increased or decreased with equal probability - approximately half of the utterances have a slower speed compared to the original set and the other half have a faster speed. Volume is handled in a similar way. All audio processing is done using Sox\footnote{\url{http://sox.sourceforge.net/}}. Different combinations of speed and volume perturbation were tested to see the combined performance impact on the development set~\cite{heymans2021mtr}. We only report on the results for the best setup on each dataset.

\subsection{Controlled environment using the LibriSpeech corpus}
\label{sec:dataset}
A controlled environment is created to develop a GAN and make certain design choices. We use the LibriSpeech corpus for this, because it is a freely available corpus with separated clean and noisy recordings. Because we are studying a resource-scarce environment, we limit our training data to 100 hours and only use the small language model (\textit{tg-small}). 

The 100-hour clean subset of the LibriSpeech corpus is used as `clean' set for GAN training and to develop a baseline model. An encoded set is created by downsampling the audio to 8kHz and applying WAV49 encoding to the clean set. After encoding, the file is upsampled back to 16kHz to train a wide-band model. This is done because we want to use a GAN as a front-end to an existing 16kHz system. The upsampling process used does not attempt to interpolate values to add high-frequency information that was lost during downsampling. Only the existing lower frequency components are retained. We also create a development and test set using the \textit{dev-clean} and \textit{test-clean} subsets of the LibriSpeech corpus. Table \ref{tab:wav49_librispeech_datasets} shows the training, development and test sets that are created for the first experiment.

\begin{table}[b!]
\centering
\caption{WAV49-encoded training, development and test sets created using the LibriSpeech \textit{train-clean-100}, \textit{dev-clean} and \textit{test-clean} sets.}
\begin{tabular}[t]{l c c c}
    \hline
    \textbf{Dataset Name}\hspace{0.25cm} & \textbf{Source Dataset} & \hspace{0.25cm}\textbf{Encoding}\hspace{0.25cm} & \textbf{Hours} \\
    \hline
    train-clean-e-100 & train-clean-100 & WAV49 & 100.6 \\
    dev-clean-e & dev-clean & WAV49 & 5.4 \\
    test-clean-e & test-clean & WAV49 & 5.4 \\
    \hline
\end{tabular}
\label{tab:wav49_librispeech_datasets}
\end{table}

In under-resourced environments, it can very expensive to collect many hours of training data. To show the ability of a Guided-GAN to work with limited training data, we create datasets ranging from 1 hour to 100 hours using the \textit{train-other-500} subset of the LibriSpeech corpus. We also encode the data with WAV49 encoding. Random speakers are selected to get a more diverse training set. The largest set (100 hours) contains 250 speakers. The datasets we use for the second experiment are shown in Table \ref{tab:ls_datasets}.

\begin{table}[b!]
\centering
\caption{LibriSpeech dataset partitions used to simulate a resource-scarce environment.}
\begin{tabular}[t]{l c c}
    \toprule
    \textbf{Dataset} & \textbf{Encoding} & \textbf{Hours} \\
    \midrule
    train-clean-100 & - & 100.6 \\
    train-clean-e-100 & WAV49 & 100.6 \\
    \midrule
    train-other-e-1 & WAV49 & 1.0 \\
    train-other-e-5 & WAV49 & 5.0 \\
    train-other-e-10 & WAV49 & 10.0 \\
    train-other-e-20 & WAV49 & 20.3 \\
    train-other-e-50 & WAV49 & 50.0 \\
    train-other-e-100 & WAV49 & 106.2 \\
    \midrule
    \textbf{Multi-style training sets} \\
    \hspace{0.25cm}train-clean-e-100-s & WAV49 & 100.6 \\
    \hspace{0.25cm}train-other-e-100-s & WAV49 & 106.2 \\
    \hspace{0.25cm}train-other-e-100-v & WAV49 & 106.2 \\
    \hspace{0.25cm}train-other-e-100-sv & WAV49 & 106.2 \\
    \midrule
    dev-other-e & WAV49 & 5.3 \\
    test-other-e & WAV49 & 5.1 \\
    \bottomrule
\end{tabular}
\label{tab:ls_datasets}
\end{table}

\section{Experiments}
In this section, we utilise our Guided-GAN approach to improve the decoding accuracy of a baseline acoustic model and compare the results to an MTR system. We first compare three different loss functions and two GAN network architectures on a WAV49-encoded LibriSpeech corpus. WAV49 encoding is used to demonstrate the ability of the GAN to improve audio features of encoded recordings. Compression and low sampling rates are typical of low-resource datasets, which is why we encoded the LibriSpeech corpus. We then reduce the amount of training data available to the GAN and demonstrate its effectiveness with limited data. We compare the GAN to an MTR system using the same training data. Finally, we evaluate the Guided-GAN on a proprietary South African Call Centre corpus to demonstrate its ability to work on real-world mismatch datasets.

\subsection{Clean WAV49-encoded LibriSpeech corpus}
\label{sec:clean_wav49_librispeech}
Our first experiment uses the LibriSpeech corpus that is encoded with WAV49 encoding. The datasets used in this experiment are shown in Table \ref{tab:wav49_librispeech_datasets}. A baseline acoustic model is trained using the \textit{train-clean-100} subset. We improve this model by using the encoded training dataset (\textit{train-clean-e-100}) and retraining a new system. We also train an MTR system with added speed perturbation.

A GAN is trained using the \textit{train-clean-100} subset as `clean' set and the \textit{train-clean-e-100} set as `mismatched' set. We compare the three GAN loss functions and two network architectures described in Sections \ref{sec:loss_functions} and \ref{sec:network_architectures}, respectively. The first GAN we train uses the baseline \textit{train-clean-100} acoustic model for guidance. This model was trained using 16kHz unencoded audio which requires the GAN to compensate for sampling rate differences and encoding. Another GAN is trained using the \textit{train-clean-e-100} acoustic model, which already used 8kHz encoded audio for training. In this case, the GAN does not have to compensate for sampling rate differences or encoding because the acoustic model was already trained on the same type of audio. 

Table \ref{tab:results_gan_librispeech_encoded} shows the results on the \textit{dev-clean-e} and \textit{test-clean-e} subsets. The top section shows the results for the baseline, encoded and MTR system that used speed perturbation (\textit{train-clean-e-100 + s})\footnote{This notation means two datasets were used: \textit{train-clean-e-100} and \textit{train-clean-e-100-s}. We follow this notation in the rest of this paper.}. The GANs are evaluated before and after fine-tuning (indicated by `FT' in the column heading). The GANs using the encoder-decoder network architecture are labeled with `E-D' and the networks using the fully-convolutional architecture with `F-C'. The middle section shows the results for the GANs using different loss functions and the \textit{train-clean-100} acoustic model for guidance. The GANs in the bottom section use the \textit{train-clean-e-100} acoustic model for guidance.

All four GANs using the baseline acoustic model significantly improved the performance. Fine-tuning further reduced the WER to be comparable with MTR. The SN-GAN loss function performed better than the NS-GAN and WGAN-GP for this task. The two network architectures, encoder-decoder and fully-convolutional, performed very similar. Both GAN models using the \textit{train-clean-e-100} acoustic model without fine-tuning performs worse than the model used to train them. When fine-tuning this model, the performance is slightly better than the MTR system. 

The two GAN models using the fully-convolutional architecture are also evaluated on the \textit{test-clean-e} set to confirm the results. The results are similar to those on the development set, although the average WER of the MTR model was slightly lower than the \textit{train-clean-100} GAN. These results confirm that the Guided-GAN is effective to recover most of the lost performance due to the reduced sampling rate and WAV49 encoding. The results of the GAN is comparable to MTR.

\begin{table}[t!]
\centering
\caption{WER results of Guided-GANs compared to baseline, encoded and MTR models on \textit{dev/test-clean-e}. The GANs using the encoder-decoder network architecture are labeled with an `E-D', and the GANs using the fully-convolutional network architecture are labeled with an `F-C'. Average WER and standard error is shown over three seeds. Best results (in bold) are comparable.}
\begin{tabular}[t]{l c c c c}
    \toprule
    \textbf{Model} & \textbf{Dev WER} & \textbf{Dev WER (FT)} & \textbf{Test WER} & \textbf{Test WER (FT)} \\
    \midrule
    train-clean-100 & 19.32 $\pm$ 0.11 & - & 19.15 $\pm$ 0.11 & - \\
    train-clean-e-100 & 11.02 $\pm$ 0.03 & - & 11.10 $\pm$ 0.02 & - \\
    train-clean-e-100 + s (MTR) & \textbf{10.93 $\pm$ 0.08} & - & \textbf{11.00 $\pm$ 0.07} & - \\
    \midrule
    \textbf{train-clean-100} \\
    \hspace{0.25cm}NS-GAN (E-D) & 13.45 $\pm$ 0.12 & 11.14 $\pm$ 0.05 & - & - \\
    \hspace{0.25cm}WGAN-GP (E-D) & 13.52 $\pm$ 0.04 & 11.14 $\pm$ 0.02 & - & - \\
    \hspace{0.25cm}SN-GAN (E-D) & 12.80 $\pm$ 0.01 & \textbf{11.04 $\pm$ 0.03} & - & - \\
    \hspace{0.25cm}SN-GAN (F-C) & 13.41 $\pm$ 0.04 & \textbf{10.93 $\pm$ 0.03} & 13.59 $\pm$ 0.03 & \textbf{11.04 $\pm$ 0.02} \\
    \midrule
    \textbf{train-clean-e-100} \\
    \hspace{0.25cm}SN-GAN (E-D) & 11.29 $\pm$ 0.04 & \textbf{10.81 $\pm$ 0.04} & - & - \\
    \hspace{0.25cm}SN-GAN (F-C) & 11.36 $\pm$ 0.02 & \textbf{10.79 $\pm$ 0.03} & 11.58 $\pm$ 0.11 & \textbf{10.94 $\pm$ 0.02} \\
    \bottomrule
\end{tabular}
\label{tab:results_gan_librispeech_encoded}
\end{table}

\subsection{Simulated resource-scarce environment using the LibriSpeech corpus}
We use the \textit{train-clean-100} baseline model trained in Section \ref{sec:clean_wav49_librispeech} to train Guided-GANs using the noisy datasets shown in Table \ref{tab:ls_datasets}. For each GAN, random utterances from the \textit{train-clean-100} set is used as the `clean' features for training. The `mismatched' features, used as input for the generator, are selected from the corresponding \textit{train-other-e-100} set. 

The WER of the clean baseline model on the standard \textit{test-clean} set is 9.18\%, and increases to 42.52\% when it is evaluated on a noisy WAV49 encoded set (\textit{test-other-e}). Noise and encoding is highly detrimental to the performance of this model. This large generalisation gap is due to a mismatched training and test set. The two major differences between the training and testing data is that the test set is encoded using WAV49 encoding and contains various forms of noise (additive noise, poor quality recordings, etc.).

We improve this model by adding a Guided-GAN front-end. We train our GAN using different amounts of training data and also fine-tune the acoustic models each time. Table \ref{tab:wer_results_time} shows the WER on the \textit{dev/test-other-e} sets. Using only one hour of noisy data for GAN training, improved the baseline by 20.2\% relative WER. Adding additional noisy data reduced the WER further, however, the returns were diminishing.

The Guided-GAN can significantly improve a clean baseline acoustic model using very little matched training data. In resource-scarce environments, where it is sometimes difficult to collect many hours of transcribed data, a Guided-GAN can be a useful tool to adapt an existing system. An additional advantage of the architecture is that the GAN is quite fast to train, especially on small datasets. 

\begin{table}[tb!]
\centering
\caption{WER results on \textit{dev/test-other-e} (noisy, encoded data) using different amounts of training data. Average WER is shown over three seeds.}
\begin{tabular}[t]{l c c}
    \toprule
    \textbf{Model} & \textbf{Dev WER}& \textbf{Test WER} \\
    \midrule
    train-clean-100 & 39.42 & 42.52 \\
    Guided-GAN with train-other-e-1 & 31.79 & 33.92 \\
    Guided-GAN with train-other-e-5 & 29.84 & 31.26 \\
    Guided-GAN with train-other-e-10 & 28.90 & 30.67 \\
    Guided-GAN with train-other-e-20 & 27.53 & 29.47 \\
    Guided-GAN with train-other-e-50 & 26.61 & 28.29 \\
    Guided-GAN with train-other-e-100 & \textbf{25.79} & \textbf{27.57} \\
    \bottomrule
\end{tabular}
\label{tab:wer_results_time}
\end{table}

Different GANs using the \textit{train-other-e-100} set are compared to MTR systems in Table \ref{tab:wer_results_gan}. The top section shows the results for the two baseline models trained with the \textit{train-clean-100} and \textit{train-clean-e-100} sets, also used by the GANs for training. The second section shows the results for models that improve the baseline results by using MTR and/or encoded noisy training data. We show results for the MTR models that use speed and volume perturbation in a single set (\textit{train-other-e-100 + sv}) and two sets separately (\textit{train-other-e-100 + s + v}). Two GANs are trained using the \textit{train-clean-100} and \textit{train-clean-e-100} acoustic models for guidance. We fine-tune the acoustic models (indicated with an `FT' label) with the generator operating on the features. We also fine-tune the acoustic model using the MTR datasets to further improve the WER. 

\begin{table}[b!]
\centering
\caption{WER results on \textit{dev/test-other-e} of baseline, MTR and GAN models using different combinations of training datasets. Fine-tuning is indicated with an `FT' label. Average WER and standard error is shown over three seeds. Best results (in bold) are comparable.}
\begin{tabular}[t]{l c c}
    \toprule
    \textbf{Model} & \textbf{dev-other-e} & \textbf{test-other-e} \\
    \midrule
    \textbf{Baseline models} \\
    \hspace{0.25cm}train-clean-100 & 39.42 $\pm$ 0.57 & 42.52 $\pm$ 0.57 \\
    \hspace{0.25cm}train-clean-e-100 & 29.01 $\pm$ 0.10 & 31.16 $\pm$ 0.16 \\
    \midrule
    \textbf{Improved models} \\
    \hspace{0.25cm}train-clean-e-100 + s & 28.57 $\pm$ 0.04 & 30.67 $\pm$ 0.09 \\
    \hspace{0.25cm}train-other-e-100 & \textbf{25.32 $\pm$ 0.08} & \textbf{26.87 $\pm$ 0.11} \\
    \hspace{0.25cm}train-other-e-100 + sv & \textbf{25.19 $\pm$ 0.08} & \textbf{26.61 $\pm$ 0.10} \\
    \hspace{0.25cm}train-other-e-100 + s + v & \textbf{24.99 $\pm$ 0.01} & \textbf{26.49 $\pm$ 0.12} \\
    \midrule
    \textbf{train-clean-100} \\
    \hspace{0.25cm}Guided-GAN & 31.25 $\pm$ 0.07 & 33.71 $\pm$ 0.03 \\
    \hspace{0.25cm}Guided-GAN + FT & 25.79 $\pm$ 0.10 & 27.57 $\pm$ 0.07 \\
    \midrule
    \textbf{train-clean-e-100} \\
    \hspace{0.25cm}Guided-GAN & 27.82 $\pm$ 0.20 & 29.29 $\pm$ 0.02 \\
    \hspace{0.25cm}Guided-GAN + FT & \textbf{25.27 $\pm$ 0.01} & \textbf{27.02 $\pm$ 0.02} \\
    \hspace{0.25cm}Guided-GAN + MTR FT (s + v) & \textbf{24.98 $\pm$ 0.05} & \textbf{26.70 $\pm$ 0.07} \\
    \bottomrule
\end{tabular}
\label{tab:wer_results_gan}
\end{table}

Applying the GAN to the \textit{train-clean-100} and \textit{train-clean-e-100} models improved the WER of the baseline acoustic models by 20.7\% and 6.0\%, respectively (relative improvements). Fine-tuning the acoustic models using the \textit{train-other-e-100} set further reduced the WER by 18.2\% and 7.7\%, respectively. Fine-tuning the \textit{train-clean-e-100} acoustic model using the speed- and volume-perturbed datasets yielded the best performance of all GANs. The results of the Guided-GANs are very comparable to MTR in terms of WER. In Section \ref{sec:gan_vs_mtr_training_time}, we compare the training time of the models to see the true benefit of the GAN.

\subsection{Comparing training times}
\label{sec:gan_vs_mtr_training_time}
We compare the training time of Guided-GANs using different amounts of training data to MTR systems. Table \ref{tab:training_time_compare} shows the training time in minutes and test WER of each model.

\begin{table}[t!]
\centering
\caption{Training time comparison of a Guided-GAN, baseline and MTR systems on a 100-hour training set (\textit{train-other-e-100}). Fine-tuning is indicated with an `FT' label.}
\begin{tabular}[t]{l c c c}
    \toprule
    \textbf{Model} & \textbf{Test WER} & \textbf{Training Time} \\
    \midrule
    train-clean-100 & 42.52 & 267 \\
    train-clean-e-100 & 31.16 & 271 \\
    train-other-e-100 & 26.87 & 288 \\
    train-other-e-100 + sv & 26.61 & 583 \\
    train-other-e-100 + s + v & 26.49 & 875 \\
    \midrule
    \textbf{train-clean-100} \\
    \hspace{0.25cm}Guided-GAN with train-other-e-1 + FT & 33.92 & 10 (4 + 6) \\
    \hspace{0.25cm}Guided-GAN with train-other-e-5 + FT & 31.26 & 35 (16 + 19) \\
    \hspace{0.25cm}Guided-GAN with train-other-e-10 + FT & 30.67 & 52 (30 + 22) \\
    \hspace{0.25cm}Guided-GAN with train-other-e-20 + FT & 29.47 & 96 (46 + 50) \\
    \hspace{0.25cm}Guided-GAN with train-other-e-50 + FT & 28.29 & 140 (60 + 80) \\
    \hspace{0.25cm}Guided-GAN with train-other-e-100 + FT & 27.57 & 195 (80 + 115) \\
    \midrule
    \textbf{train-clean-e-100} \\
    \hspace{0.25cm}Guided-GAN & 29.29 & 80 \\
    \hspace{0.25cm}Guided-GAN + FT & 27.02 & 195 (80 + 115) \\
    \hspace{0.25cm}Guided-GAN + MTR FT (s + v) & 26.70 & 425 (80 + 345) \\
    \bottomrule
\end{tabular}
\label{tab:training_time_compare}
\end{table}

Training a GAN requires more memory than training a standard MLP acoustic model because the GAN uses three networks during training (generator, discriminator and acoustic model). Although it uses more memory, the training time is less than training an MTR model because the GAN converges much faster than an acoustic model. The GAN also uses much larger batch sizes during training, which also speeds it up significantly. 

The training time is measured using a desktop computer running an Intel i7 8700k CPU with an Asus RTX 3080 GPU. The system has 32 GB RAM running at 3 200 MHz. The baseline and MTR models were trained for 24 epochs. The GANs using the 100-hour training set were trained for 20 epochs. The GANs using less data were trained for more epochs. All networks converged before the end of training was reached. We always use early stopping on the validation SeER to select the best GAN network, but for this experiment, we compare the training time until the end of training. 

Training a Guided-GAN using the fully-convolutional architecture and fine-tuning is about three times faster than training an MTR model using two datasets and four and a half times faster than one using three sets. The MTR systems perform very similarly to the GAN in terms of WER. Fine-tuning the GAN with the MTR datasets is still two times faster than training an MTR model using the same sets. The GAN is also extremely efficient with limited training data. A relative WER improvement of 20.2\% on the \textit{train-clean-100} model is achieved with only 10 minutes of GAN training, including fine-tuning time. With 10 hours of training data, the relative WER improvement increases to 27.9\% with a GAN trained in 52 minutes.

This shows that the Guided-GAN is not only a viable replacement for MTR in resource-scarce environments, it is also an efficient technique to improve a good baseline system on mismatched data. The only requirement is that a small, transcribed dataset with conditions similar to the test set is available.

\subsection{Call centre dataset}
To evaluate the effectiveness of a Guided-GAN on a real-world resource-scarce dataset, we use a proprietary South African Call Centre (SACC) corpus. A baseline system is trained using 48.8 hours of audio recorded in a specific call centre in South Africa (\textit{Baseline train}). The lexicon and language model are specific to this call centre and are used by all models. We use the same MLP acoustic model and training procedure described in Section \ref{sec:acoustic_model}. This baseline model achieves 28.41\% WER on a development set (\textit{Baseline dev}) containing calls from the same call centre not included in the training data. This is considered matched data, since the calls originate from the same call centre.

Small datasets from eight different call centres are available as mismatched sets, shown in Table \ref{tab:small_call_centre_datasets_info}. The data in these sets are extremely limited, ranging from 11 minutes to 75 minutes. For each call centre, a train, development and test set are available. The datasets have different sampling rates, bitrates and encoding types, shown in Table \ref{tab:small_call_centre_datasets_info} for each set. We select three call centre datasets to train individual GAN models on and train another GAN using all eight datasets (labeled `\textit{Mixed centres}'). This mixed set is used to train a GAN with more conditions (from all call centres) to show the ability of the GAN to generalise to a number of different environments. The data between call centres differ in terms of: recording environments, noise conditions, dialects and accents, jargon, noise levels, encoding, etc.

The \textit{Baseline train} set is used as the `clean' or `matched' set for GAN training and the training set of the new call centres as `mismatched' set. The hyperparameters are optimised to minimise the WER on the development set. After the best hyperparameters are selected, a new GAN is trained with the training and development set to expand the training data. This is done because the datasets are extremely small and additional data is very beneficial.

\begin{table}[b!]
\centering
\caption{Data subsets in the SACC corpus with the total duration in minutes.}
\begin{tabular}[t]{l c c c c c}
    \toprule
    \textbf{Dataset} & \textbf{Sampling Rate} & \textbf{Bitrate} & \textbf{Encoding} & \textbf{Calls} & \textbf{Minutes} \\
    \midrule
    Baseline train & 8kHz & 128kbps & - & 727 & 2 928.3 \\
    Baseline dev & 8kHz & 128kbps & - & 101 & 425.5 \\
    \midrule
    Call centre 1 train & 16kHz & 256kbps & - & 8 & 38.8 \\
    Call centre 1 dev & 16kHz & 256kbps & - & 6 & 20.7 \\
    Call centre 1 test & 16kHz & 256kbps & - & 5 & 15.5 \\
    \midrule
    Call centre 2 train & 8kHz & 64kbps & A-Law & 12 & 43.2 \\
    Call centre 2 dev & 8kHz & 64kbps & A-Law & 3 & 12.8 \\
    Call centre 2 test & 8kHz & 64kbps & A-Law & 3 & 15.3 \\
    \midrule
    Call centre 3 train & 8kHz & 256kbps & - & 2 & 6.6 \\
    Call centre 3 dev & 8kHz & 256kbps & - & 1 & 2.9 \\
    Call centre 3 test & 8kHz & 256kbps & - & 1 & 1.9 \\
    \midrule
    Mixed centres train & 8-16kHz & 64-256kbps & mixed & 50 & 351.2 \\
    Mixed centres dev & 8-16kHz & 64-256kbps & mixed & 19 & 145.7 \\
    Mixed centres test & 8-16kHz & 64-256kbps & mixed & 26 & 188.9 \\
    \bottomrule
\end{tabular}
\label{tab:small_call_centre_datasets_info}
\end{table}

The WER results for the GANs compared to the baseline model are shown in Table \ref{tab:gan_call_centre_datasets}. The performance of the baseline model on the new mismatched datasets are extremely poor. All WERs on the test sets are above 50\% for the baseline model. Applying the Guided-GAN as a front-end to the baseline model consistently improves the results on all sets. The improvements on the development set ranged from 5.45\% to 24.74\%. The best test set improvement was found on Call centre 3, with only 9.5 minutes of training data, the relative WER improvement was 19.70\%.

\begin{table}[t!]
\centering
\caption{WER results of baseline and GANs on the SACC \textit{dev/test} sets. The relative WER improvement is shown in the right column.}
\begin{tabular}[t]{l c c c c}
    \toprule
    \textbf{Dataset} & \textbf{Baseline} & \textbf{GAN} & \textbf{Improvement} \\
    \midrule
    \textbf{Development set} \\
    \hspace{0.25cm}Call centre 1 & 56.58 & 45.67 & 19.28\% \\
    \hspace{0.25cm}Call centre 2 & 48.67 & 36.63 & 24.74\% \\
    \hspace{0.25cm}Call centre 3 & 70.66 & 66.81 & 5.45\% \\
    \hspace{0.25cm}Mixed centres & 61.68 & 56.11 & 9.03\% \\
    \midrule
    \textbf{Test set} \\
    \hspace{0.25cm}Call centre 1 & 55.11 & 48.77 & 11.50\% \\
    \hspace{0.25cm}Call centre 2 & 59.05 & 51.40 & 12.96\% \\
    \hspace{0.25cm}Call centre 3 & 52.38 & 42.06 & 19.70\% \\
    \hspace{0.25cm}Mixed centres & 60.63 & 55.17 & 9.01\% \\
    \bottomrule
\end{tabular}
\label{tab:gan_call_centre_datasets}
\end{table}

\section{Analysis}
\label{sec:gan_improvement_analysis}
In this section, we investigate the improvements made by a Guided-GAN. We compare the errors made between the baseline and the GAN in Section \ref{sec:error_comparison}, and visualise the improvements on a frame level in Section \ref{sec:feature_transformation}.

\subsection{Error comparison}
\label{sec:error_comparison}
We compare the errors made by the \textit{train-clean-100} baseline and Guided-GAN by using the \textit{train-clean-100} acoustic model. Table \ref{tab:error_compare} shows a summary of the errors made by the two models. The GAN reduced the number of insertions, deletions and substitutions, which resulted in a lower WER. On the frame level, the GAN improved the SeER by 14.9\% relatively. This shows that a Guided-GAN improves the acoustic model on all types of errors and not just a specific type. 

\begin{table}[htb!]
\centering
\caption{Comparison of errors made by the \textit{train-clean-100} baseline and Guided-GAN trained by using the \textit{train-clean-100} acoustic model on the LibriSpeech \textit{dev-clean-e} set.}
\begin{tabular}[t]{l c c}
    \toprule
    \textbf{Metric} & \textbf{Baseline}& \textbf{Guided-GAN} \\
    \midrule
    Insertions & 984 & 746 \\
    Deletions & 1 317 & 616 \\
    Substitutions & 8 228 & 4 569 \\
    Total errors & 10 529 & 5 931 \\
    Total words & 54 402 & 54 402 \\
    WER & 19.35\% & 10.90\% \\
    SeER & 48.90\% & 41.60\% \\
    \bottomrule
\end{tabular}
\label{tab:error_compare}
\end{table}

\subsection{Feature transformation}
\label{sec:feature_transformation}
\begin{figure}[h]
    \centering
    \includegraphics[width=0.95\textwidth]{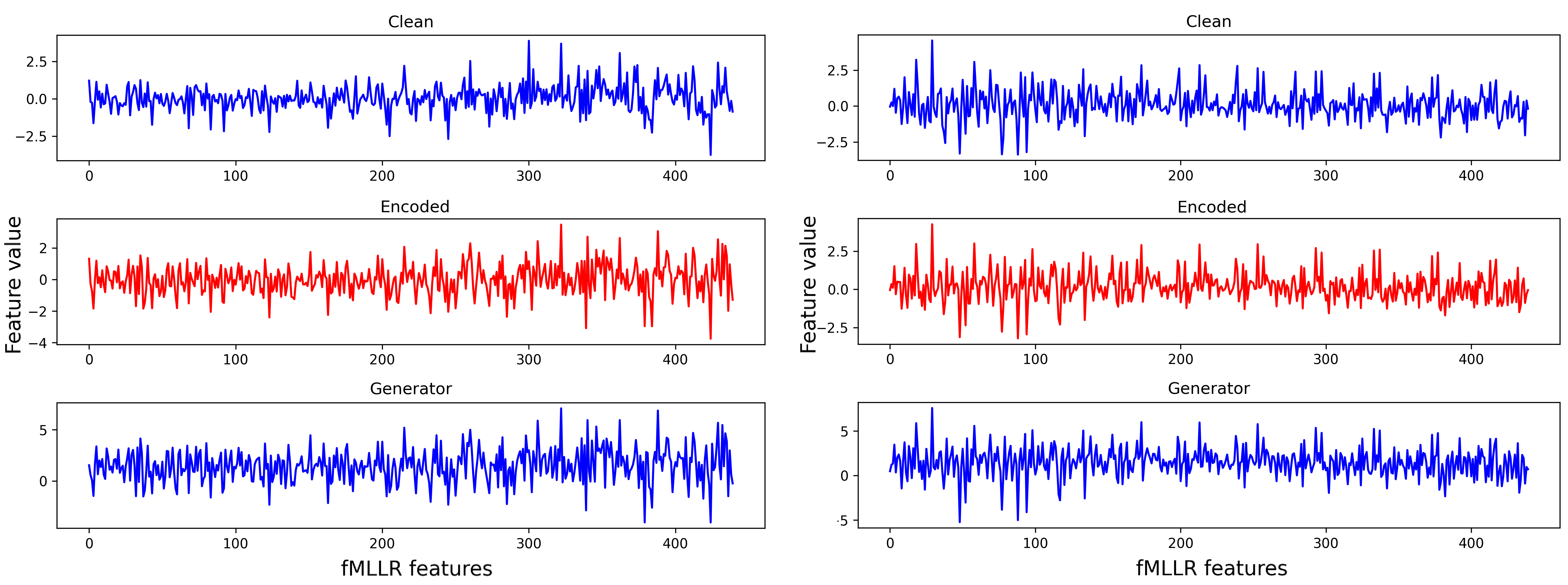} 
    \caption{Features of clean, encoded and generated samples of two different utterances.}
    \label{fig:generated_features}
\end{figure}

To visualise the transformation that a Guided-GAN makes on a frame level, we plot the fMLLR features of two clean, encoded and generated samples. We use the features of a clean sample and the WAV49-encoded version of the same sample. We show two examples in Figure \ref{fig:generated_features}. The encoded and generated features on the left were classified incorrectly by the acoustic model. In the figure on the right, the encoded features were classified incorrectly, but the generated features were classified correctly.

On average, the encoded features correlate slightly better with the clean features than the generated features with the clean features. This indicates that the generator does not aim to recover the original features, but instead changes the features in some way so that the acoustic model better classifies them. This is further confirmed by using a trained generator on another acoustic model that was not used for guidance (different initialisation seed used for the acoustic model). If the features were closer to the original features, we would expect that the new acoustic model will also be improved, but this is not the case. Instead, the performance is worse than when using the model without a GAN. The Guided-GAN can only be used to improve the model that was used for guidance during training because the generator only learns how to improve the classification accuracy of that specific model.

\section{Discussion}
Feature transformation using a GAN can significantly reduce the WER of a baseline model when there is a mismatch between the training and testing audio. Similar improvements are observed as when using MTR, at less cost: training a GAN front-end takes much less time than retraining an ASR system using MTR. The GAN can be used with or without fine-tuning, with a fine-tuned acoustic model producing improved accuracy but at an additional computational cost. Fine-tuning the GAN with MTR datasets provided further improvements.

The acoustic model used to calculate the acoustic model loss for the generator during training is very important. The \textit{train-clean-100} baseline never used encoded audio for training, which meant that it had no idea how to handle encoding. This requires the generator to learn a transformation that can compensate for sampling rate differences and remove encoding artifacts from the features. Since WAV49 is a lossy encoding, the task is very difficult, and the large gain observed from the \textit{train-clean-100} baseline when only the GAN was added (14.95\% absolute WER) is very promising. 

This work is aimed primarily at under-resourced environments where computational capabilities are limited. The GAN is fast to train and can be plugged into existing systems without other modifications. This is very useful if a strong baseline model has already been developed, but there is a desire to use it on new data from a mismatched environment. Using MTR or other enhancement techniques that require manual creation of a noisy or better matched dataset are limited to the conditions that can be artificially created. It is extremely difficult to accurately create a noisy dataset that is well matched to a new set of conditions. The performance of a model using a dataset that was manually created is far worse than one with perfectly matched conditions~\cite{heymans2021mtr}. When a small dataset with matched conditions is available, current GAN techniques cannot be used to reduce the mismatch, because a corresponding clean version does not exist (they require both the clean and noisy version to minimise the L1 or mean squared error loss). The Guided-GAN does not have this limitation because it utilises a baseline acoustic model to provide guidance during training instead of an L1 or mean squared error loss.

\section{Conclusion}
Our GAN framework can significantly improve a baseline ASR system that is evaluated on mismatched data. The Guided-GAN solves many of the shortcomings of existing GAN or enhancement systems. Current techniques require parallel clean and noisy training data. It is extremely difficult to accurately reproduce a training set that is matched to unknown noise conditions. The developed GAN can utilise a small matched dataset to adapt an existing baseline system trained on good quality data to also perform well on the mismatched data.

 The practical relevance of a Guided-GAN is realised when a strong commercial-grade ASR system is required to decode new mismatched data. In commercial applications, it is very expensive to retrain or tweak an ASR system that has already been optimised for a given set of conditions. The strong ASR system can be re-used on the new dataset by using a Guided-GAN. A small training set from the target environment can be collected and manually transcribed without much difficulty. The Guided-GAN can then use this training set to adapt the strong ASR system to improve its performance on the new dataset.

\section*{Acknowledgement}
We would like to thank the South African Centre for High Performance Computing (CHPC) who made their facilities available for some of these experiments.

\bibliographystyle{IEEEtran}
\balance
\bibliography{mybib}

\appendix{}
\renewcommand\thefigure{\thesection.\arabic{figure}} 
\setcounter{figure}{0}

\section{Appendix: Correlation between WER and SeER}
\label{apx:correlation}
\begin{figure}[ht!]
    \centering
    \includegraphics[width=0.6\textwidth]{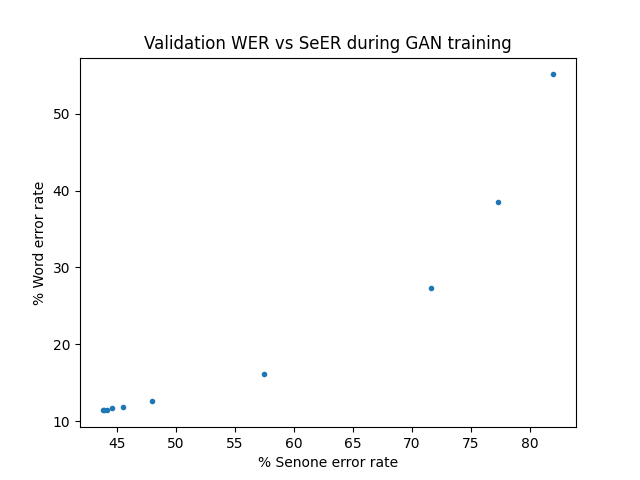}
    \caption{WER versus SeER on development set measured with trained acoustic model during GAN training.}
    \label{fig:wer_seer_gan_training}
\end{figure}

\begin{figure}[ht!]
    \centering
    \includegraphics[width=0.6\textwidth]{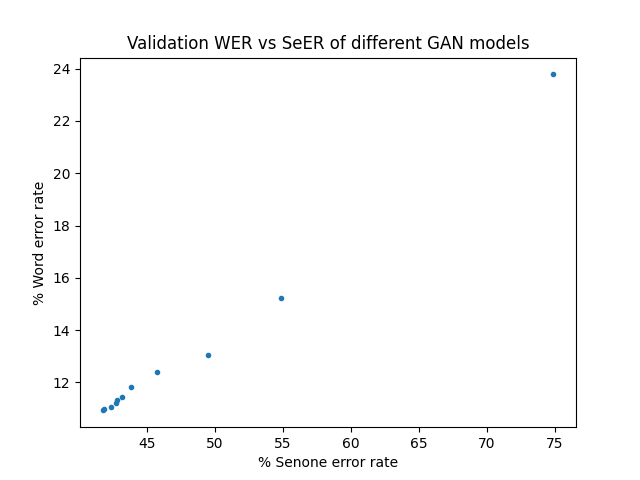}
    \caption{WER versus SeER on development set for 11 different GAN models after training.}
    \label{fig:wer_seer}
\end{figure}

To confirm the correlation between WER and SeER, we measure both during GAN training on the LibriSpeech corpus. Ten points are selected where we measure the SeER and perform decoding using the MLP acoustic model. The data points were selected ranging from the early stages in training up to a much later point where the network has converged. We do not fine-tune the acoustic model in this experiment. Figure \ref{fig:wer_seer_gan_training} shows the WER versus SeER during GAN training. A strong correlation exists between WER and SeER. We also evaluated eleven different types of GAN models after training. A graph of WER versus SeER for these models after training is shown in Figure \ref{fig:wer_seer}. The models in this graph used different network architectures, hyperparameters and loss functions. The correlation also holds here, despite the setups being very different from one another. 

\end{document}